\documentstyle[twocolumn,floats,aps,prl,epsfig]{revtex}

\begin{document}
\draft

\twocolumn[\hsize\textwidth\columnwidth\hsize\csname@twocolumnfalse\endcsname%

\title{Double-dot charge transport in Si single electron/hole
transistors}

\author{L.~P. Rokhinson, L.~J. Guo$^{a)}$, S.~Y. Chou and D.~C. Tsui}

\address{Department of Electrical Engineering, Princeton University,
Princeton, NJ 08544}

\date{To appear in \apl on March 20, 2000}

\maketitle

\begin{abstract}
We studied transport through ultra-small Si quantum dot transistors
fabricated from silicon-on-insulator wafers. At high temperatures, 4
K $<T<$ 100 K, the devices show single-electron or single-hole
transport through the lithographically defined dot. At $T<4$ K,
current through the devices is characterized by multidot transport.
From the analysis of the transport in samples with double-dot
characteristics, we conclude that extra dots are formed inside the
thermally grown gate oxide which surrounds the lithographically
defined dot.
\end{abstract}

\vskip2pc]

Recent advances in miniaturization of Si metal-oxide-semiconductor
field-effect transistors (MOSFETs) brought to light several issues
related to the electrical transport in Si nanostructures. At low
temperatures and low source-drain bias Si nanostructures do not
follow regular MOSFET transconductance characteristics but show
rather complex behavior, suggesting transport through
multiply-connected dots.  Even in devices with no intentionally
defined dots (like Si quantum
wires\cite{nakajima95,ishikuro96,hiramoto97,smith97} or point
contacts\cite{ishikuro97}) Coulomb blockade oscillations were
reported. In the case of quantum wires, formation of tunneling
barriers is usually attributed to fluctuations of the thickness of
the wire or of the gate oxide.  However, formation of a dot in point
contact samples is not quite consistent with such explanation.
Recently in an elegant experiment with both $n^+$ and $p^+$
source/drain connected to the same Si point contact Ishikuro and
Hiramoto\cite{ishikuro99} have shown that the confining potential in
unintentionally created dots is similar for both holes and electrons.
However, there is no clear picture where and how these dots are
formed.

In this work we analyze the low temperature transport through an
ultra-small lithographically defined Si quantum dots.  While at high
temperature 4 K $<T<$ 100 K we observe single-electron tunneling
through the lithographically defined dot, at $T<4$ K transport is
found to be typical for a multi-dot system.  We restrict ourselves to
the analysis of samples with double-dot transport characteristics.
From the data we extract electrostatic characteristics of both the
lithographically defined and the extra dots.  Remarkably, transport
in some samples cannot be described by tunneling through two dots
connected in sequence but rather reflects tunneling through dots
connected in parallel to both source and drain.  Taking into account
the geometry of the samples we conclude that extra dots should be
formed within the gate oxide. Transport in p- and n-type samples are
similar, suggesting that the origin of the confining potential for
electrons and holes in these extra dots is the same.

%
%
\begin{figure}[tb]
\epsfig{file=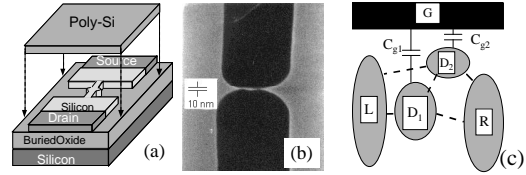,width=3.25in}
\vspace{-1.8in}
\caption{(a) Schematic of the device structure, (b) SEM micrograph of
a device, and (c) schematic view of two dots D$_1$ and D$_2$
connected to source and drain contacts L and R. G represents a gate
electrode and C$_{g1}$ and C$_{g2}$ are gate capacitances. Dashed
lines represent possible tunneling barriers.}
\label{scheme}
\end{figure}

The samples are metal-oxide-semiconductor field-effect transistors
(MOSFETs) fabricated from a silicon-on-insulator (SOI) wafer. The top
silicon layer is patterned by an electron-beam lithography to form a
small dot connected to wide source and drain regions, see schematic
in Fig.~\ref{scheme}a. Next, the buried oxide is etched beneath the
dot transforming it into a free-standing bridge. Subsequently, 40 or
50 nm of oxide is thermally grown which further reduces the size of
the dot. Poly-silicon gate is deposited over the bridge with the dot
as well as over the adjacent regions of the source and drain. It is
important to note that in this type of devices the gate not only
controls the potential of the dot but also changes the dot-source and
dot-drain barriers. Finally, the uncovered regions of the source and
drain are n--type or p--type doped. More details on samples
preparation can be found in Ref.~\cite{leobandung95}. Totally, about
30 hole and electron samples have been studied. Here we present data
from two samples with hole (H5A) and electron (E5-7) field-induced
channels.

An SEM investigation of test samples, Fig.~\ref{scheme}b, reveals
that the lithographically defined dot in the Si bridge is 10-40 nm in
diameter and the distance between narrow regions of the bridge is
$\sim$70 nm. Taking into account the oxide thickness we estimate the
gate capacitance to be 0.8--1.5 aF.

%
%
\begin{figure}[tb]
\psfig{file=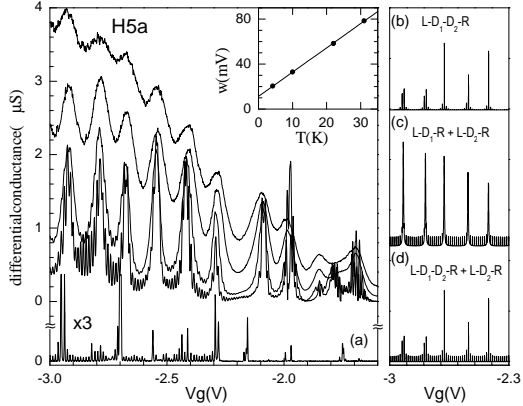,width=3.25in}
\vspace{-1.5in}
\caption{(a) Differential conductance in the hole quantum dot sample
H5A is shown as a function of the gate voltage $V_g$ for $T=31$, 22,
10, 4.2 and 0.3 K (from top to bottom). The trace at the lowest
temperature 0.3 K has been taken in a separate cooldown. In the inset
peak width $w$ vs $T$ is plotted for peaks between $-3.0<V_g<-2.2$ V
($w$ is defined in the text). (b)-(d) Modeling of the total
conductance at $T=0.3$ K assuming that the two dots are connected (b)
in series, (c) in parallel, or (d) mixed.}
\label{h5a-t-dep}
\end{figure}

\begin{figure}[tb]
\epsfig{file=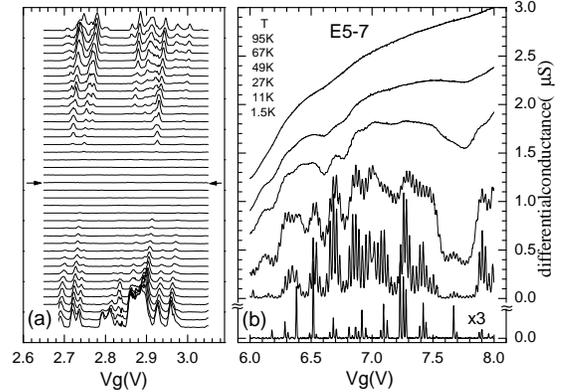,width=3.25in}
\vspace{-3.1in}
\caption{Differential conductance in an electron quantum dot sample
E5-7 is plotted as a function of the gate voltage $V_g$ for (a)
different dc source-drain bias $V_b$ and (b) different temperatures.
In (a) each curve is measured at different $V_b$ from -20 mV (bottom
curve) to 20 mV (top curve) at $T=1.5$ K. Arrows indicate the curve
with $V_b=0$. All curves are offset by 0.5 $\mu$S. Data in (b) is
taken at zero bias. The excitation voltage is 100 $\mu$V.}
\label{seq1}
\end{figure}

In most of our samples (with both n-- and p--channel) we see clear
Coulomb blockade oscillations with a period $\Delta V_{g1}=100-160$
mV up to $\sim$100 K. A typical charge addition spectra is plotted in
Fig.~\ref{h5a-t-dep} and Fig~\ref{seq1} for samples H5A and E5-7. In
H5A the spectrum is almost periodic as a function of the gate voltage
$V_g$ at $T>4$ K with the period $\Delta V_{g1}\approx 130$ mV.
Assuming that each peak corresponds to an addition of one hole into
the dot we calculate the gate capacitance $C_{g1}=e/\Delta
V_{g1}=1.2$ aF, which is within the error bars for the capacitance
estimated from the sample geometry. The lineshape of an individual
peak can be described\cite{kulik75,beenakker91} by $G\propto
\cosh^{-2}[(V_g-V_g^i)/2.5\alpha k_B T]$, where $V_g^i$ is the peak
position and coefficient $\alpha=C_{total}/eC_g$ relates the change
in the $V_g$ to the shift of the energy levels in the dot relative to
the Fermi energy in the contacts. This expression is valid if both
coupling to the leads $\Gamma$ and single-particle level spacing
$\Delta E$ are small: $\Gamma,\Delta E \ll k_B T \ll e^2/C_{total}$.
We fit the data for H5A with $\sum_{i} \cosh^{-2}[(V_g-V_g^i)/w]$ in
the range -3.0 V $<V_g<-2.2$ V and the extracted $w$ is plotted in
the inset in Fig.~\ref{h5a-t-dep}.  From the linear fit $w=11.3+2.2T$
[mV] we find the coefficient $\alpha=10$ [mV/meV], thus the Coulomb
energy is $\approx 13$ meV and the total capacitance $C_{total}=12.3$
aF. The main contribution to $C_{total}$ comes from dot-to-lead
capacitances (an estimated self-capacitance is a few aF). The
extrapolated value of $w$ at zero temperature provides an estimate
for the level broadening $\Gamma\approx1$ meV.

At $T<4$ K oscillations with another period, much smaller than
$\Delta V_{g1}$, appear as a function of $V_g$. The small period is
in the range $\Delta V_{g2}=8-25$ mV in different devices ($\Delta
V_{g2}=11.8$ mV for the sample in Fig.~\ref{h5a-t-dep}). This small
period is due to a single-hole tunneling through a second dot and the
corresponding gate capacitance $C_{g2}=e/\Delta V_{g2}=6-20$ aF.
However, there is no intentionally defined second dot in our devices.
Below we first analyze the experimental results and then discuss
where the second dot can be formed.

%
%

At low temperatures and small gate voltages (close to the turn-on of
the device at high temperatures) current is either totally
suppressed, as in E5-7 at  $V_g<3.5$ V, Fig.~\ref{seq1}a, or there
are sharp peaks with no apparent periodicity, as in H5A at $V_g>-2.3$
V, Fig.~\ref{h5a-t-dep}. Both suppression of the current and
``stochastic Coulomb blockade''\cite{ruzin92} are typical signatures
of tunneling through two sequentially connected dots. The non-zero
conductance can be restored either by raising the temperature
(Fig.~\ref{h5a-t-dep}) or by increasing the source-drain bias $V_b$
(Fig.~\ref{seq1}a). In both cases, $G$ is modulated with $\Delta
V_{g1}$ and $\Delta V_{g2}$, consistent with sequential tunneling. We
conclude that in these regime the two dots are connected in series
L-D$_1$-D$_2$-R (see schematic  in Fig.~\ref{scheme}c).

At larger gate voltages ($V_g>6$ V for E5-7 and $V_g<-2.3$ V for H5A)
current is not suppressed even at the lowest temperatures. However,
the $G$ pattern is different in the H5A and E5-7 samples. In H5A, the
oscillations with $\Delta V_{g2}$ have approximately the same
amplitude (except for the sharp peaks which are separated by
approximately $\Delta V_{g1}$), while in E5-7 the amplitude of the
fast oscillations is modulated by $\Delta V_{g1}$. Also, the
dependence of the amplitude of the fast modulations on the average
conductance $<G>$ is different: in H5A the amplitude is almost
$<G>$-independent, while in E5-7 it is larger for larger $<G>$.

Non-vanishing periodic conductance at low temperatures requires that
the transport is governed by the Coulomb blockade through only one
dot D$_2$. That can be achieved either if both barriers between the
contacts and the D$_2$ become transparent enough to allow substantial
tunneling or if the strong coupling between the main dot D$_1$ and
one of the leads results in a non-vanishing density of states in the
dot at $T=0$. If we neglect coupling between the dots, in the former
case the total conductance is approximately the sum of two
conductances, $G_{parallel}\approx G_1+G_2$, where $G_1$ is
conductance through the main dot L-D$_1$-R and $G_2$ is conductance
through the second dot L-D$_2$-R. This case is modeled in
Fig~\ref{h5a-t-dep}c using experimentally determined parameters of
sample H5A. From the analysis of high-temperature transport we found
that the zero-temperature broadening of D$_1$ peaks
$\alpha\Gamma\approx10$ mV $\approx\Delta V_{g2}\ll \Delta
V_{g1}=130$ mV and that $G$ should be exponentially suppressed
between D$_1$ peaks at $T=0.3$ K if the dots are connected in series
L-D$_1$-D$_2$-R, Fig~\ref{h5a-t-dep}b. The best description of the
low temperature transport at -3.0 V $ <V_g<-2.3$ V in H5A is achieved
if we assume that there are two conducting paths in parallel: through
the extra dot L-D$_2$-R and through both dots together
L-D$_1$-D$_2$-R, Fig~\ref{h5a-t-dep}d.

In the latter case, the dots are connected in series L-D$_1$-D$_2$-R.
At high $V_g$ the barrier between L and D$_1$ is reduced giving rise
to a large level broadening $\Gamma$. The total conductance is
$G_{series}\approx G_{BW} G_2/(G_{BW}+G_2)$, where $G_2$ is the
Coulomb blockade conductance through D$_2$ alone and
$G_{BW}={{2e^2}\over{h}} \Gamma^2/(\Gamma^2+\delta E^2)$ is the
Breit-Wigner conductance through D$_1$ and $\delta
E=(V_g-V_g^i)/\alpha$. In this case $G_{series}$ is following
$G_{BW}$ and is modulated by $G_2$. Moreover, if we assume that the
amplitude of $G_2$ is not a strong function of $V_g$, the amplitude
of $G_{series}$ modulation will be a function of $G_{BW}$, namely the
larger $G_{BW}$ the larger the amplitude of the modulation of the
total conductance. This model of two dots in series with one being
strongly coupled to the leads is in qualitative agreement with the
data from sample E5-7.

\begin{figure}[tb]
\epsfig{file=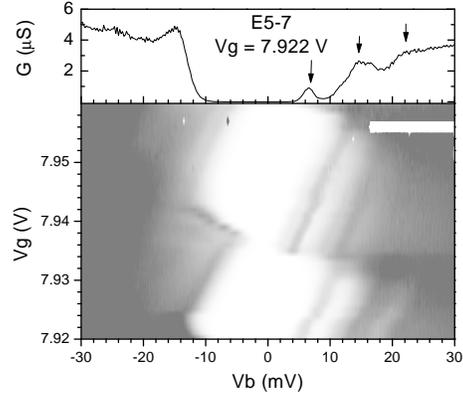,width=3.25in}
\vspace{0in}
\caption{Differential conductance on a gray scale as a function of
both $V_g$ and $V_b$. A single trace at $V_g=7.922$ is shown at the
top. Arrows indicate onset of the tunneling of 1,2 and 3 electrons
simultaneously, as discussed in the text.}
\label{E57-gray}
\end{figure}

Non-equilibrium transport through E5-7 is shown in
Fig.~\ref{E57-gray} with a single $G$ vs. $V_b$ trace at a fixed
$V_g$ shown at the top of the figure. White diamond-shaped Coulomb
blockade regions are clearly seen on the gray-scale plot. Peaks in
$G$ at positive bias are due to asymmetry in the tunneling
barriers\cite{su92}: at negative biases tunneling to the dot is
slower than tunneling off the dot and only one extra electron
occupies the dot at any given time, thus only one peak, corresponding
to the onset of the current, is observed (we have not seen any
features due to the size quantization, which is not surprising if we
take into account the large number of electrons in this dot). At
positive biases current is limited by the time the electron spends in
the dot before it tunnels out. In this regime an extra step in the
I-V characteristic (and a corresponding peak in its derivative $G$)
is observed every time one more electron can tunnel into the dot.
These peaks, marked with arrows, are separated by the charging energy
$U_c=e\Delta V_b=8$ meV.

%
%
%
%

Electrostatic parameters of the D$_2$ dot can be readily extracted
from Fig.~\ref{E57-gray}. The source, drain and gate capacitances are
8.5, 2.7 and 6.4 aF and the corresponding charging energy is
$\approx9$ meV. The charging energy of $\approx11$ meV is obtained by
analyzing Fermi-Dirac broadening of the conductance peaks as a
function of temperature and the period of oscillations. The fact that
it requires the application of $V_b=10$ mV to lift the Coulomb
blockade means that in the Coulomb blockade regime all the bias is
applied across the second dot, consistent with large conductance
through D$_1$.

Where does the second dot reside? One possibility is that the silicon
bridge, containing the lithographically defined dot, breaks up at low
temperatures as a result of the depletion due to variations of the
bridge thickness and fluctuations in the thickness of the gate oxide,
or due to the field induced by ionized impurities. However, in this
case $C_{g2}$ should be less than $C_{g1}$.  In fact, if we assume
that the  thickness of the thermally grown oxide is uniform, the gate
capacitance of the largest possible dot in the channel cannot be
larger than 1.5 aF. Also, if at low temperatures the main dot would
split into two or more dots we should see the change in the period of
the large oscillations\cite{waugh95}, inconsistent with our
observations.

Another possibility is that the dot is formed in the contact region
adjacent to the bridge.  Given that the oxide thickness is 40 nm, the
second dot diameter should be $\approx 100$ nm. We measured two
devices which have 30 nm wide and 500 nm long channels, fabricated
using the same technique as the dot devices. Both samples show
regular MOSFET characteristics down to 50 mK. Thus, it is unlikely
that a dot is formed in the wide contact regions of the device. Even
if such a dot was formed occasionally in some device by, for example,
randomly distributed impurities, it is unlikely that dots of
approximately the same size would be formed in all samples. Another
argument against such a scenario is that if the second dot is formed
inside one of the contact regions, it cannot be coupled to the other
contact to provide a parallel conduction channel, as in sample H5A.

Thus, the second dot should reside within the gate oxide, which
surrounds the lithographically defined dot. Some traps can create
confining potential in both conduction and valence bands, for example
P$_b$ center has energy levels at $E_c-0.3$ eV and $E_v+0.3$ eV.
Several samples show a hysteresis during large gate voltage scans
accompanied by sudden switching. This behavior can be attributed to
the charging-discharging of traps in the oxide. If such a trap
happens to be in a tunneling distance from both the lithographically
defined dot and a contact, or the trap is extended from one contact
to the other, it may appear as a second dot in the conductance.

To summarize our results, we performed an extensive study of a large
number of Si quantum dots. We found that all devices show multi-dot
transport characteristics at low temperatures. From the data analysis
we arrived at the conclusion that at least double-dot behavior is
caused not by the depletion of the silicon channel but by additional
transport through traps within the oxide.

We acknowledge the support from ARO, ONR and DARPA.

\end{document}